\documentclass{article}

\newcommand{\be}{\begin{equation}}
\newcommand{\ee}{\end{equation}}

\def\bea{\begin{eqnarray}}
\def\eea{\end{eqnarray}}
\def\bean{\begin{eqnarray*}}
\def\eean{\end{eqnarray*}}

\newcommand{\barr}{\begin{array}}
\newcommand{\earr}{\end{array}}

\newcommand{\bed}{\begin{displaymath}}
\newcommand{\eed}{\end{displaymath}}
\newcommand{\bal}{\begin{array}{ll}}
\newcommand{\eal}{\end{array}}

\def\bvec#1{\raise1.5ex\hbox{$\rightarrow$}\mkern-16.5mu #1}

\def\jrn#1#2#3#4{{#1} {\bf #2} (#4) #3}
\def\PRL{Phys. Rev. Lett.}
\def\PLB{Phys. Lett. B}
\def\PRD{Phys. Rev. D}
\def\NPB{Nucl. Phys. B}

\begin{document}

\pagestyle{empty}

\rightline{UFIFT-HEP-03/14}

\vspace*{1.5cm}

\begin{center}

\LARGE{Correlated Hierarchy, Dirac Masses  and Large Mixing Angles\\[20mm]}

\large{Aseshkrishna Datta\footnote{e-mail: datta@phys.ufl.edu}, 
Fu-Sin~Ling\footnote{e-mail: fsling@phys.ufl.edu} and
Pierre Ramond\footnote{e-mail: ramond@phys.ufl.edu}\\[8mm]}

\it{Institute for Fundamental Theory\\
Department of Physics, University of Florida,\\ 
Gainesville, FL, 32611, USA\\[15mm]}

\large{\rm{Abstract}} \\[7mm]

\end{center}

\begin{center}

\begin{minipage}[h]{14cm}
We introduce a new parametrization of the MNS lepton mixing matrix which  
separates the hierarchical Grand Unified relations among quarks and leptons. 
We argue that one large angle stems from the charged leptons, the other 
from the seesaw structure of the neutral lepton mass matrix. 
We show how two large mixing angles can arise naturally provided 
there are special requirements on the Dirac ($\Delta I_{\rm w}=1/2$) and Majorana 
($\Delta I_{\rm w}=0$) masses. One possibility is a correlated hierarchy 
between them, the other is that the $\Delta I_{\rm w}=0$ Majorana mass 
has a specific texture; it is Dirac-like for two of 
the three families.
\end{minipage}

\end{center}

\newpage

\pagestyle{plain}

\section{Introduction}

The most  compelling scheme for physics beyond the Standard Model revolves 
around the idea of Grand Unification between quarks and leptons.  
This pattern is evident in terms of the fermion quantum numbers,  
in view of the unification of the weak and strong coupling constants at 
high energy, and more recently with the discovery of light neutrino masses. 
The well-documented hierarchy of quarks and charged lepton masses  seems to 
apply as well in the neutrino sector, although not decisively since their 
masses are inferred solely from interference effects, and there could be 
hyperfine splittings in their spectrum. 
Yet, recent experiments~(\cite{SKatm}~-~\cite{K2K}) on neutrino  
mixings show  qualitatively different  behavior for quarks and leptons. 
Quark mixings seems to be approximately the same for charge $2/3$ and charge 
$-1/3$ quarks, resulting in a CKM matrix that is approximately the unit matrix, 
save for Cabibbo angle effects. 
On the other hand, two of the three mixing angles in the MNS lepton  
mixing matrix are ``large",  and only  one is small. This implies that 
the mechanisms that determine the lepton mass eigenstates are different for 
charged and neutral leptons. 

There are differences between the fermion masses even at the level of the 
Standard Model. The quark and charged lepton masses are Dirac-like and break electroweak 
symmetry along $\Delta I_{\rm w}=1/2$, but the neutrino masses can be either 
Dirac-like with the same $\Delta I_{\rm w}=1/2$ structure or Majorana with 
$\Delta I_{\rm w}=1$; in the latter case  qualitative differences may be expected.

As Grand Unification relates the Dirac matrices of quarks and leptons, 
one may expect some if not all of the quark hierarchy to be mirrored  
in the lepton sector~\cite{SU5SO10}.  
However, $SO(10)$ unification brings about the seesaw mechanism for 
generating neutrino masses through  a new fermion with a large 
$\Delta I_{\rm w}=0$ Majorana mass~\cite{SEESAW}. 

In previous works~\cite{HRR, FM}, based on $SO(10)$ and on the 
Froggatt-Nielsen (FN) approach~\cite{FN}
to masses and mixings, we anticipated,  before the SuperKamiokande data, 
one large mixing angle between  the second and third family neutrinos in 
the MNS matrix, and two small mixing angles from the first to the second 
and third family neutrinos. 
The recent data on solar neutrinos proved otherwise: the MNS matrix 
contains two large mixing angles, and only one small one.  

The purpose of this note is to investigate in simple terms the possible 
origin of this unexpected second large angle, without having to ditch 
cherished notions from Grand Unification. 
In particular, we would like to analyze how large mixing angles are 
related to a hierarchical structure in the mass matrix.

We expect a generic ({\it i.e.} without any hierarchical structure) 
$3\times 3$  matrix to be diagonalized by rotations 
with three angles which are not particularly small.  
One can also imagine quite easily a matrix that is diagonalized by one 
large mixing and two small ones, by assuming inequalities between matrix 
elements of one row and one column with the remaining ones. 
Similarly, if all the off-diagonal matrix elements are small enough,
the mixing matrix will nearly the the unit matrix and all three 
mixing angles come out small. 
However, the CHOOZ constraint~\cite{CHOOZ} indicates that 
only one mixing angle in the MNS matrix is small. 
This is not possible to arrange in the hierarchical structure alone, 
and fine-tuning between various matrix elements is needed. 

The MNS matrix is the overlap of two unitary matrices, 
one from the charged lepton sector, the other from the neutral sector.  
The search for a generic mechanism suggests, in view of the previous remarks, 
that  one large mixing angle comes from the diagonalization of  
the charged lepton masses, the other from that of  the neutral lepton masses. 
We argue below that the large angle found in atmospheric neutrino oscillations 
comes from the $\Delta I_{\rm w}=1/2$ charged lepton matrix, with a {\it modicum} 
of assumptions. The large angle recently determined in $\nu_e$ oscillations then 
comes from the special form of the seesaw mechanism, either through a correlated 
hierarchy of the Dirac and $\Delta I_{\rm w}=0$ Majorana  masses, or through a 
Dirac-like texture~\cite{RRR} for two of the three the right-handed neutrino 
mass matrix. We conclude with some remarks on how these ideas can be incorporated 
into a theory of the Froggatt-Nielsen type.  

\section{Neutrino Masses Just Beyond the Standard Model}

We begin by reminding the reader how masses and mixings arise in the Standard 
Model, stressing their properties in terms of electroweak quantum numbers. 
The usual $\Delta I_{\rm w}=1/2$ electroweak breaking generates quarks and charged 
leptons masses and mixing angles through Yukawa couplings.  
In the quark sectors, it gives rise to Dirac masses through  

\be
{\cal M}^{(2/3)}_{}~={\cal U}^{}_{2/3}\,
\pmatrix{m^{}_u&0&0\cr 0&m^{}_c&0\cr 0&0&m_t^{}}
\,{\cal V}^{\dagger}_{2/3}\ .
\ee

\be
{\cal M}^{(-1/3)}_{}~=~{\cal U}^{}_{-1/3}\,
\pmatrix{m^{}_d&0&0\cr 0&m^{}_s&0\cr 0&0&m_b^{}}
\,{\cal V}^{\dagger}_{-1/3}\ ,
\ee
where $\cal U$ and $\cal V$ are unitary matrices which diagonalize them. 
Not all these matrices are directly observable: of these four, only the CKM 
quark mixing matrix 
\be
{\cal U}^{}_{CKM}~=~{\cal U}^{\dagger}_{2/3}\,{\cal U}^{}_{-1/3}\ ,
\ee
is observable, and the right-handed matrices ${\cal V}^{}_{2/3}$, 
${\cal V}^{}_{-1/3}$ are not physical. 

Similarly in the charged lepton sector, 

\be
{\cal M}^{(-1)}_{}~=~{\cal U}^{}_{-1}\,
\pmatrix{m^{}_e&0&0\cr 0&m^{}_\mu&0\cr 0&0&m_\tau^{}}
\,{\cal V}^{\dagger}_{-1}\ .
\ee
In the original version of the Standard Model, neither unitary matrix is 
observable, and all three neutrinos are massless. 
All  mass matrices in the quarks and charged lepton sectors arise from 
the same $\Delta\,I^{}_{\rm w}=1/2$ electroweak breaking mechanism.

To go beyond the Standard Model, we add one right-handed neutrino for each 
family, and the same electroweak breaking generates a Dirac mass for the neutrinos
\be
{\cal M}^{(0)}_{Dirac}~=~{\cal U}^{}_{0}\,{\cal D}_0^{}\,
{\cal V}^{\dagger}_{0}~=~{\cal U}^{}_{0}\,
\pmatrix{m^{}_1&0&0\cr 0&m^{}_2&0\cr 0&0&m_3^{}}\,{\cal V}^{\dagger}_{0}\ ,
\label{LEPTON}
\ee
where $m_i$ are the neutrino {\it Dirac} masses,   {\bf not} to be confused 
with their physical masses (unless there is no seesaw).  
This matrix respects total lepton number $L_e+L_\mu+L_\tau$, 
but violates the relative lepton numbers $L_e-L_\mu$ and $L_\mu-L_\tau$. 

To account for  small neutrino masses, it is customary to assign large 
Majorana masses to these three right-handed neutrinos.
Let ${\cal M}^{(0)}_{Majorana}$ be the Majorana mass matrix with  entries of 
order $M$, assumed to be much larger than the electroweak breaking scale. 
This matrix can be markedly different from the others as it arises not from 
electroweak breaking,  but from some unknown $\Delta I_{\rm w}^{}=0$ sector. 
After seesaw ($M\gg m$), the neutral mass matrix is  
\be
{\cal M}^{(0)}_{Seesaw}~=~{\cal M}^{(0)}_{Dirac}\,
\frac{1}{{\cal M}^{(0)}_{Majorana}}\,{\cal M}^{(0)\,T}_{Dirac}\ ,
\ee
where $T$ stands for the transpose. 
Inserting the diagonalization~(\ref{LEPTON}), we find

\bea{\cal M}^{(0)}_{Seesaw}&=&{\cal U}^{}_{0}\,
{\cal D}_0^{}\,{\cal V}^{\dagger}_{0}\,\frac{1}{{\cal M}^{(0)}_{Majorana}}\,
{\cal V}^{*}_{0}\,{\cal D}_0^{}\,{\cal U}^{T}_{0}\ ,\cr &=&{\cal U}^{}_{0}\,
{\cal C}\,{\cal U}^{T}_{0}\ ,\nonumber\eea
where $\cal C$ is the central matrix

\be{\cal C}~\equiv~{\cal D}_0^{}\,{\cal V}^{\dagger}_{0}\,
\frac{1}{{\cal M}^{(0)}_{Majorana}}\,{\cal V}^{*}_{0}\,{\cal D}_0^{}\ ,\ee
which is diagonalized  by means of a unitary matrix ${\cal F}$ 

\be{\cal C}~=~{\cal F}\,{\cal D}^{}_\nu\,{\cal F}^T_{}\ ,\ee 
where ${\cal D}_\nu$ is the diagonal matrix 
\be
{\cal D}_\nu^{}~=~
\pmatrix{m^{}_{\nu_1}&0&0\cr 0&m^{}_{\nu_2}&0\cr 0&0&m_{\nu_3}^{}}\ ,
\ee
with the {\it physical} neutrino masses as entries. 
This enables us to write the observable 
MNS lepton mixing matrix  in the suggestive form

\be{\cal U}^{}_{MNS}~=~ {\cal U}^{\dagger}_{-1}\,{\cal U}^{}_{0}\,{\cal F}\ ,\ee
where  $\cal F$ diagonalizes the central matrix of the seesaw. 
It is similar in form to the CKM quark mixing matrix in that it contains 
the overlap between two unitary matrices that  diagonalize the  
$\Delta I_{\rm w}=1/2$ mass matrices, but it also contains a totally new 
matrix which comes from the right-handed $\Delta I_{\rm w}=0$ neutrino masses. 

This  parametrization, which separates the effect of the $\Delta I_{\rm w}=1/2$ 
Dirac masses from that of the $\Delta I_{\rm w}=0$ Majorana masses, 
makes it particularly convenient to discuss the intuition coming from 
Grand Unified Theories. 
With the simplest Higgs structures, these theories typically 
relate the $\Delta I_{\rm w}=1/2$ mass matrices for quarks and leptons 
\be SU(5)~~~:~~~~~~~~{\cal M}^{(-1/3)}_{}~\sim~{\cal M}^{(-1)\,T}_{}\ .\ee
\be SO(10):~~~~~~~~{\cal M}^{(2/3)}_{}~\sim~{\cal M}^{(0)}_{Dirac}\ .\ee
These imply similar hierarchies between the CKM and MNS matrix, 
should ${\cal F}$ be equal to one, which happens if the neutral mass matrix 
is purely Dirac-like (as in theories with bulk right-handed neutrinos). 
On the other hand, if the seesaw is operative, the effects of  
the unitary matrix ${\cal F}$ must be included. 

The $\Delta I_{\rm w}=1/2$ mass matrices in the quark and charged lepton 
sectors show hierarchy, a feature consistent with Grand Unified Theories. 
On the other hand, over the last five years, several experiments have shown 
that the MNS matrix  shares this hierarchy only partially, as it contains one 
small  and two large mixing angles. 
We would like to argue that in the context of GUTs, this may be an indication 
in favor of the seesaw, with one or both large angles in the MNS matrix 
coming from ${\cal F}$. 

However, the central matrix $\cal C$ already contains some hierarchical information 
coming from ${\cal D}_0$, the eigenvalues of the hierarchical neutral Dirac mass. 
Indeed, using Grand Unification as a rough guide, we expect a hierarchical structure

\be
m_1~\ll~ m_2~\ll~m_3\ ,\ee
but if $\cal F$ is to contain large angles, some of the matrix elements of 
the central matrix must be of similar orders of magnitude.  
This can happen through numerical accidents, or generically if  
${\cal M}^{(0)}_{Majorana}$ contains a  {\it correlated hierarchy} of its own, 
designed to offset that of the Dirac masses.  
The purpose of this paper is to find how $\cal F$ can contain large angles 
despite the expected hierarchy of the Dirac masses.  

\section{The Data}

Let us see how these theoretical expectations compare with the data. 
The combined experimental results favor a scheme with three active neutrinos.
The mixing matrix $U_{MNS}$ has two large angles 
($\theta$ and $\phi$ for atmospheric~\cite{SKatm} and
solar~\cite{SKsol, SNO}  neutrinos respectively)  
and one small one (from the CHOOZ~\cite{CHOOZ} constraint). 
This enables us to write  

\be {\cal U}^{}_{MNS}~\simeq~
\pmatrix{\cos\phi&\sin\phi&\epsilon\cr -\cos\theta\,\sin\phi& 
\cos\theta\,\cos\phi&\sin\theta\cr
\sin\theta\,\sin\phi&-\sin\theta\,\cos\phi&\cos\theta}\ ,
\ee
where $\epsilon$ is a complex number, with the limits~\cite{fogli}

\be \sin^2_{}2\theta \geq 0.85 \; (99\% \;{\rm C.L.})\ ,\qquad 
0.30 \leq \tan^2_{}\phi \leq 0.65 \; (95\% \; {\rm C.L.})\ ,\qquad 
\vert \epsilon \,\vert^2 \leq .005 \; (99.73\% \;{\rm C.L.})\  .\ee
The neutrino mass splittings are

$$\Delta m^2_\odot~=~\vert m^2_{\nu_1}-m^2_{\nu_2}\vert~\simeq~
7\times 10^{-5}_{} \; {\rm eV}^2_{}\ ,\qquad 
\Delta m^2_\oplus ~=~\vert m^2_{\nu_2}-m^2_{\nu_3}\vert~\simeq~
3\times 10^{-3}_{} \; {\rm eV}^2\ ,$$
coming from the solar and atmospheric neutrino oscillations, respectively. 
Although neutrino oscillation experiments do not fix the absolute mass
scale of the neutrinos, we now have very stringent bounds from cosmology.
The large scale structure formation and the latest measurements of the
cosmic microwave background by WMAP~\cite{WMAP} indicate that 
\be
\sum _k \, m_{\nu _k} ~\leq~ 0.71 \; {\rm eV} \; (95\% \; {\rm C.L.})
\ee

These are consistent with three possible neutrino mass patterns 

\begin{picture}(200,170)

\thicklines
\put(50,70){\line(1,0){50}}
\put(50,110){\line(1,0){50}}
\put(50,150){\line(1,0){50}}
\put(30,70){\shortstack{$\nu_1$}}
\put(30,110){\shortstack{$\nu_2$}}
\put(30,150){\shortstack{$\nu_3$}}
\put(55,45){\shortstack{Hierarchy}}
\put(30,25){\shortstack{$|m_{\nu _1}| 
\leq |m_{\nu _2}| \ll |m_{\nu _3}|$}}

\put(180,70){\line(1,0){50}}
\put(180,140){\line(1,0){50}}
\put(180,150){\line(1,0){50}}
\put(160,70){\shortstack{$\nu_3$}}
\put(160,140){\shortstack{$\nu_2$}}
\put(160,150){\shortstack{$\nu_1$}}
\put(190,45){\shortstack{Inverted}}
\put(160,25){\shortstack{$|m_{\nu _1}| 
\simeq |m_{\nu _2}| \gg |m_{\nu _3}|$}}

\put(310,110){\line(1,0){50}}
\put(310,120){\line(1,0){50}}
\put(310,130){\line(1,0){50}}
\put(290,110){\shortstack{$\nu_1$}}
\put(290,120){\shortstack{$\nu_2$}}
\put(290,130){\shortstack{$\nu_3$}}
\put(315,45){\shortstack{Hyperfine}}
\put(290,25){\shortstack{$|m_{\nu _1}| 
\simeq |m_{\nu _2}| \simeq |m_{\nu _3}|$}}

\end{picture}

We can use the data to reconstruct the neutrino matrix in all three cases, 
and look for patterns~\cite{KM}.

\begin{itemize}
\item In the hierarchy case, we neglect $m_{\nu_1}$ and $m_{\nu_2}$, 
and the Yukawa matrix looks like

\be 
\pmatrix{0&0&0\cr 0&m^{}_{\nu_3}\,\sin^2_{}\theta &m^{}_{\nu_3}\,
\sin\theta\,\cos\theta\cr
0&m^{}_{\nu_3}\,\sin\theta\,\cos\theta&m^{}_{\nu_3}\,\cos^2_{}\theta}\ ,
\ee
which suggest as a starting point a matrix with entries in the $23$ block only
\be 
\pmatrix{0&0&0\cr 0&~\times~ &~\times~\cr
0&~\times~&~\times~}\ ,\ee
where the crosses describe elements of order one. 
However, in order to obtain from this pattern the second large angle, 
this matrix must contain {\it two} zero eigenvalues, that is the sub-determinant 
must vanish. This type of matrix has been used as a starting point for the 
Froggatt-Nielsen approach, with corrections of the order of the Cabibbo angle 
in previous works~\cite{FM, me1}, 
and it can lead to a second large mixing angle only if 
the sub-determinant is very small~\cite{me1, sato, vissani}.  
If the sub-determinant is not small, non-degenerate perturbation theory applies 
and the solar neutrino mixing angle comes out small~\cite{me3}.

\item In the inverted hierarchy case, all entries of the matrix seem to be of the 
same order of magnitude, but there is a special case of great interest. 
If we take $m_{\nu_1}\approx -m_{\nu_2}$ and maximal angles, the matrix looks like

\be
\pmatrix{0&~\times~ &~\times~ \cr ~\times~ &0&0\cr ~\times~ &0&0}\ ,
\label{invstruct}
\ee
which can be used as a starting point for a Cabibbo expansion~\cite{king}. 
In this special case, the masses of the first two neutrinos follow a 
nearly Dirac pattern, with a global $L_e - L_\mu - L_\tau$ 
symmetry~\cite{babu}. 

\item Finally, in the hyperfine case where all three masses are nearly degenerate, 
there is no discernible pattern in the Yukawa matrix. 

\end{itemize}

We would like to see how to reproduce these patterns, starting from 
conservative theoretical notions, namely Grand Unified symmetries. 

\section{A Useful Approximation}

In the following we would like to argue that the large mixing angle found 
in atmospheric neutrino oscillations stems from ${\cal U}_{-1}$, the unitary 
matrix in the charged lepton sector. To that effect, consider the limit in 
which the Dirac masses have two zero eigenvalues in each charged sector, so 
that only the third family is massive.  If we extend the Wolfenstein 
parametrization of the CKM matrix to quark and charged 
lepton masses, expressing the masses of the lighter families as powers of the 
Cabibbo angle, this limit can be thought of as that 
where  the Cabibbo angle is zero.   
In this limit, we set the two quark mass matrices as

$${\cal M}^{(2/3)}_{}~=~\pmatrix{0&0&0\cr 0&0&0\cr 0&0&m_t^{}}~=~
{\cal U}^{}_{2/3}\,\pmatrix{0&0&0\cr 0&0&0\cr 0&0&m_t^{}}\,{\cal V}^{\dagger}_{2/3}\ .$$

$${\cal M}^{(-1/3)}_{}~=~\pmatrix{0&0&0\cr 0&0&0\cr 0&a&b}~=~
{\cal U}^{}_{-1/3}\,\pmatrix{0&0&0\cr 0&0&0\cr 0&0&m_b^{}}\,{\cal V}^{\dagger}_{-1/3}\ .$$
It follows that 

$${\cal U}^{}_{2/3}~=~{\cal U}^{}_{-1/3}~=~1\ ,$$
which shows that the CKM matrix 

$${\cal U}^{}_{CKM}~=~{\cal U}^{\dagger}_{2/3}\,{\cal U}^{}_{-1/3}~=~1\ ,$$
which is a  close approximation to data. 
It also suggests that the quark family mixings in the 
$\Delta I_{\rm w}^{}~=~1/2$ sector are hypercharge independent.  

Note that we have allowed for non-zero entries in the third row of the charge 
$-1/3$ mass matrix. This still yields two zero eigenvalues and does not affect 
the left-handed rotations. In fact we have encountered such a structure in our 
previous work~\cite{FM, me1, me2} based on the 
Froggatt-Nielsen formalism, where the FN charges 
of $\overline{\bf d}_2$ and $\overline{\bf d}_3$ are naturally the same, 
leading to the same degree of suppression.  

This conclusion is not limited~\cite{SYAB} to the F-N formalism, and it also appears in a 
completely different approach,  a model 
where the first two families share the same weak $SU(2)_{1+2}$ but the third 
family has a different  weak $SU(2)_3$. 
Upon diagonal breaking, these merge into the physical weak $SU(2)_{1+2+3}$. 
This pattern of symmetries  naturally arises with three copies of the 
electroweak group~\cite{BR}, broken diagonally by a tri-chiral order parameter. 
This order parameter merges the three $SU(3)$'s into one but breaks the 
three $SU(2)$ into these two. 
With only a Higgs doublet with respect to $SU(2)_3$ at tree level, 
this singles out the third row  of the charge $2/3$ and $-1/3$  
matrices with order one entries. 
In this approximation,  two whole families of quarks are massless, 
and there is no inter-family mixing. 

To find the lepton matrices, we rely on the simplest $SU(5)$ pattern 
which  gathers the left-handed anti-down quarks with the charged 
leptons in its $\overline{\bf 5}$ representation. 
This suggests we  assign the same FN charges to the left-handed 
leptons and anti-down quarks, and leads us to assume that  
the charged lepton and charge $-1/3$ mass matrices are transpose of one another

$${\cal M}^{(-1)}_{}~=~\pmatrix{0&0&0\cr 0&0&a\cr 0&0&b}~=~{\cal U}^{}_{-1}\,
\pmatrix{0&0&0\cr 0&0&0\cr 0&0&m_\tau^{}}\,{\cal V}^{\dagger}_{-1}\ ,$$
leading to 

$${\cal U}^{}_{-1}~=~\pmatrix{1&0&0\cr 0&\cos\theta&\sin\theta\cr 0&-
\sin\theta&\cos\theta}\ ,\qquad \tan\theta~=~\frac{a}{b}\ .$$
We claim this is the origin of the large mixing angle which describes the oscillations 
of atmospheric neutrinos. 

At the next level of Grand Unification, $SO(10)$, the charge $2/3$ and the 
neutral {\it Dirac} mass matrices are supposed to be related, and if one shows 
hierarchy, so does the other. 
Hence in this approximation, we set  the neutral {\it Dirac} mass matrix  

\be
{\cal M}^{(0)}_{Dirac}~=~\pmatrix{0&0&0\cr 0&0&0\cr 0&0&m}~=~{\cal U}^{}_{0}\,
\pmatrix{0&0&0\cr 0&0&0\cr 0&0&m_3^{}}\,{\cal V}^{\dagger}_{0}\ .\label{LIMIT}\ee
It follows that  ${\cal U}^{}_{0}$ is the unit matrix and the MNS matrix becomes

\be{\cal U}^{}_{MNS}~=~{\cal U}^{\dagger}_{-1}\,{\cal U}^{}_{0}\,{\cal F}~=~
\pmatrix{1&0&0\cr 0&\cos\theta&\sin\theta\cr 0&-\sin\theta&\cos\theta}\,{\cal F}\ .\ee
The unitary matrices ${\cal U}_{-1}$ and ${\cal U}_0$ stem from the 
same $\Delta\,I^{}_{\rm w}~=~1/2$ 
electroweak breaking, but ${\cal F}$ can be markedly different 
as it arises  but from the unknown  $\Delta I_{\rm w}^{}~=~0$ sector. 

Assume that in this limit  the Majorana matrix has no zero eigenvalue. 
Since the  neutral Dirac mass matrix is of the form (\ref{LIMIT}), 
the  seesaw  neutrino mass  matrix is just
$${\cal M}^{(0)}_{Seesaw}~=~\frac{m^2}{M}\,\pmatrix{0&0&0\cr 0&0&0\cr 0&0&1}\ ,$$
so that  ${\cal F}=1$ and the MNS matrix is simply
\be{\cal U}^{}_{MNS}~=~\pmatrix{1&0&0\cr 0&\cos\theta&\sin\theta\cr 0&-
\sin\theta&\cos\theta}\ .\ee
If the Majorana matrix has  zero eigenvalues, the seesaw may not apply, 
depending on the relative alignment of the zeros in the Dirac and Majorana matrices, 
so one can have both Dirac and Majorana matrices for the neutrinos.  

With  two massless neutrinos, the family mixing described by $\theta$ does not 
specify which of the massless neutrinos  the third one mixes into, and ${\cal F}$ 
cannot change the structure of the MNS matrix.  
It is only by breaking the degeneracy  that one can identify the flavor into 
which the third neutrino mixes. With this caveat in mind,  the MNS matrix remains 
the same as above. We therefore think  it natural to expect one large lepton mixing 
angle in the limit with two massless families, even though there is no quark 
mixing (${\cal U}_{CKM}=1$). 

To lift the degeneracy, degenerate perturbation theory must be used, so that 
the form of the perturbation  determines the  mixing angle between the  two 
hitherto massless families. 
In the quark sector, that mixing is Cabibbo suppressed, and in the lepton sector 
it is of order one. 
As our analysis shows, this difference is naturally accommodated if there is a 
seesaw mechanism: the large mixing must be due to the $\cal F$ matrix. 
Given the form of ${\cal U}_{-1}$, there are two possibilities for ${\cal F}$

\be
{\cal F}~=~\pmatrix{\cos\phi&\sin\phi&0\cr -\sin\phi&\cos\phi&0\cr 0&0&1}\ ,\qquad 
{\cal F}~=~\pmatrix{\cos\phi&\sin\phi&0\cr  0&0&1\cr-\sin\phi&\cos\phi&0}\ ,\ee
obtained by permuting the second and third entries. 
The second case is the same as the first provided that one reflects $\theta$ 
about $\pi/4$. 
Either way, the mixing between the  third family and the others is naturally 
small  since non-degenerate perturbation theory applies: the smallness of the 
$13$ element  is the only remnant of the quark and charged lepton hierarchy  
in the MNS matrix!

We conclude that the second large mixing angle between the first two family 
neutrinos must arise when the Cabibbo perturbation is turned on.  
It is the purpose of the next section to see how this can come about. 

\section{Correlated Hierarchy}

In this section, we examine in some detail how the seesaw's central matrix  
can have elements of the same order of magnitude, thereby cancelling the Dirac 
hierarchy in the numerator with that in the Majorana denominator. 
At first sight, it  implies  relations between two hitherto unrelated sectors of  
the Dirac ($\Delta I_{\rm w}=1/2$) and Majorana ($\Delta I_{\rm w}=0$) masses, 
and  speaks further for Grand Unification: it suggests that the mechanism which 
creates hierarchies acts in some sense on the whole Grand-Unified  structure. 
We present a detailed analysis only for the ($2\times 2$) case, and simply state 
the results for the realistic ($3\times 3$) case.  

\subsection{The ($2\times 2$) Case}

Since  the data can be fitted with only one large mixing angle in $\cal F$, 
this case is quite instructive on its own.  
To see how a correlated hierarchy can arise generically when the neutral 
Dirac matrix is hierarchical, we expand  ${\cal D}_0$   in  the Cabibbo 
angle {\it \` a la}  Wolfenstein,
\be
{\cal D}^{}_0~=~m\pmatrix{a\,\lambda^\alpha_{}&0\cr 0&1}\ ,\ee
with $a$ of order one and $\alpha>0$, and  set 
\be{\cal V}^{\dagger}_{0}\,
\frac{1}{{\cal M}^{(0)}_{Majorana}}\,{\cal V}^{*}_{0}~=~
\pmatrix{c&s\cr -s&c}\,\pmatrix{\frac{1}{M_1}&0\cr 0&\frac{1}{M_2}}\,
\pmatrix{c&-s\cr s&c}\ ,\ee
where $c=\cos\zeta$, $s=\sin\zeta$.  The central matrix is then

\be
{\cal C}~=~\pmatrix{(\frac{c^2}{M_1}+\frac{s^2}{M_2})\,a^2_{}\,
\lambda^{2\alpha}_{}&(\frac{c\,s}{M_1}-\frac{c\,s}{M_2})\,a\,\lambda^\alpha_{}\cr
(\frac{c\,s}{M_1}-\frac{c\,s}{M_2})\,a\,\lambda^\alpha_{}&
(\frac{s^2}{M_1}+\frac{c^2}{M_2})}\ .
\label{central22}
\ee
For $\cal F$ to contain generically a large angle, the diagonal entries 
in $\cal C$ must be at most of the same order of magnitude as the 
off-diagonal element. For  ${\cal C}$ given by Eq.~(\ref{central22}), 
two different cases lead to a large angle.

\subsubsection{Case A}

In the first situation, assume all the entries in ${\cal C}$ are of the
same order of magnitude
\be
{\cal C}_{11} ~\sim~ {\cal C}_{22} ~\sim~ {\cal C}_{12} \ .
\ee 
For this to happen, $\zeta$ must be small, such that
\be s~=~b\,\lambda^\alpha_{}+\cdots\ ,\qquad c~=~1-\cdots\ ,\ee
and the Majorana masses must show hierarchy, 

\be
\frac{M_1}{M_2}~\sim~\lambda^{2\beta}_{}\ ,\ee
which must be less than or equal to $\lambda^{2\alpha}$, in order to get 
${\cal C}_{22}$ to be of the same order as the other elements. 
If $\beta>\alpha$, the  matrix $\cal C$ reduces to

$$\lambda^{2\alpha}\,\frac{m^2}{M^{}_1}\,
\pmatrix{{a^2}&{a\,b}\cr {a\,b}&{b^2}}\ ,$$
which produces {\it unsuppressed mixing} between the first two species. 
If the Majorana hierarchy is extreme ($\beta\gg \alpha$), there is one zero 
eigenvalue to lowest 
order and a {\it de facto} mass hierarchy, but with a large mixing angle. 
If the Majorana mass hierarchy is less pronounced ( $\alpha \simeq \beta$), 
the determinant no longer vanishes at lowest order, and the masses of the 
two lightest neutrinos are of the same order

\be
m^{}_{\nu_1}~\sim~m^{}_{\nu_2}~\sim~\frac{m^2}{M_1}\,\lambda^{2\beta}\ .\ee 
We see that it is possible to obtain large mixing as long  as the right-handed 
masses are also hierarchical~\cite{SMIR}.  
Moreover the Majorana hierarchy must be twice as severe as the Dirac hierarchy. 
The result is

\be
{\cal F}~=~\pmatrix{\cos\phi&\sin\phi \cr -\sin\phi&\cos\phi}\ ,\qquad 
\tan\phi~=~\frac{a}{b}\ ,\ee
to lowest order in $\lambda$. 
Because the $(3 \times 3)$ case can always be reduced to the $(2 \times 2)$
case by setting to zero the mixing angles with the third family,
we see that it is natural in this approach to expect the MNS matrix 
to contain two large angles, $\theta$ and $\phi$, while the $U_{e3}$ 
entry is naturally Cabibbo suppressed. 

\subsubsection{Case B}

The second situation which generically gives rise to a large angle is when
the diagonal entries in ${\cal C}$ are small compared to the off-diagonal
ones
\be
{\cal C}_{11}\, , \,{\cal C}_{22} ~\prec~ {\cal C}_{12}\ .
\ee
This implies a correlation between the angle $\zeta$ and the Majorana
hierarchy
\be
\tan^2 \zeta ~=~ -\frac{M_1}{M_2}\left(1+{\cal O}(\lambda^\beta)\right)
\label{sitB}
\ee
with $\beta > 0$ and
\be
\lambda^\alpha ~\prec~ s ~\preceq~ \lambda^{\alpha-\beta}
\ee
When the inequality is strict, the matrix ${\cal C}$ reduces to
$$\frac{\lambda^\alpha~m^2}{\sqrt{-M_1 M_2}}\,
\pmatrix{0&a\cr a&0}\ ,$$
which produces {\it maximal} mixing between two light neutrinos
with the same absolute mass. In particular, if the right-handed
neutrinos are given a large Dirac mass, it corresponds to
$M_1 = -M_2$ and $\zeta \simeq \pi/4$, therefore naturally ensuring
a large mixing angle among the light neutrinos.

When $s\sim\lambda^{\alpha-\beta}$, the diagonal elements of the central matrix 
cannot be neglected, and the mixing angle is somewhat less than 
maximal~\footnote{ It is amusing to note that if one diagonal element gets 
filled as much as the off-diagonal element, the mixing angle is determined 
by the celebrated golden mean which corresponds to a mixing angle of $31^o$ close 
to the experimental best-fit value, as pleasing to the eye as it may be to Nature?}.

Unlike case A, the angle $\zeta$ in case B can be larger than $\lambda^\alpha$, 
as long as it is related to the Majorana hierarchy
through Eq.~(\ref{sitB}). Hence we again have one large mixing angle, but it may 
be naturally close to maximal

 \be
{\cal F}~=~\pmatrix{\cos\phi&\sin\phi\cr -\sin\phi&\cos\phi}\ ,\qquad 
\tan\phi~\simeq~\frac{\pi}{4}\ ,\ee
just by having, to lowest order in $\lambda$, the first two right-handed neutrinos 
as Dirac partners!

\subsection{The ($3\times 3$) Case}

In the realistic ($3\times 3$) case, the analysis of the central matrix is more involved. 
As we have stated in the introduction, it is not {\it generic} to expect  a 
$(3\times 3)$ matrix to be diagonalized by one small and two large  angles. 
This does not mean it is impossible, rather it implies subtle relations among 
its matrix elements.  It is much easier to expect three, one, or no large mixing angles.  
Below, we present the results of  systematic estimates of the orders of 
magnitude of the central matrix elements and possible cancellations along the 
lines of the previous section. 

To set our notation, the hierarchical Dirac matrix eigenvalues are 
\be
{\cal D}^{}_0~\sim~m\,\pmatrix{\lambda_{}^{\alpha_1+\alpha_2}&0&0\cr 0&
\lambda_{}^{\alpha_1}&0\cr 0&0&1}\ ,\ee
ignoring prefactors. We write the unitary matrix in the central matrix as the 
product of three rotations 
$$V_0~=~{\cal R}^{}_{12}\,{\cal R}^{}_{13}\,{\cal R}^{}_{23}\ ,$$
neglecting possible phases, and where ${\cal R}_{ij}$ denotes the rotation 
in the $i-j$ plane with small angle $\theta _{ij}$, so that 
$\sin \theta _{ij} \equiv s_{ij} \simeq \theta _{ij}$ and
$\cos \theta _{ij} \simeq 1$; and take ${\cal M}^{(0)}_{Majorana}$ to be
diagonal with masses $M_i$.
To facilitate the analysis, we further assume that 
\be
s^{}_{13}~\sim~ s^{}_{12}\,s^{}_{23}
\label{rel13}
\ee
are of the same order of magnitude in $\lambda$, as for the CKM matrix. 
This relation is natural when the right-handed neutrino Majorana hierarchy
is not inverted, {\it i.e.}
\be
M_1 \leq M_2 \leq M_3
\ee

\subsubsection{Case A}

This first case corresponds to a correlated hierarchy between the
Dirac and the Majorana sector, which will lead to a neutrino mass pattern
with a normal hierarchy and one large mixing angle in ${\cal F}$.
The analysis of the orders of magnitudes of the resulting  matrix elements is tedious 
although remarkably straightforward. 
To maintain the goodwill of the reader, we only present the results:

\begin{itemize}

\item If the large angle is in the $1-2$ block of $\cal F$, the structure
of the central matrix obeys
\be
{\cal C}_{11} ~\sim~ {\cal C}_{12} ~\sim~ {\cal C}_{22} ~\prec~ {\cal C}_{33}
\ee
\be
{\cal C}_{13}^2\, , \,{\cal C}_{23}^2 ~\preceq~ 
{\cal C}_{11} \cdot~ {\cal C}_{33} 
\ee
We obtain three solutions which all satisfy 
\be
\frac{M_1}{M_2} ~\preceq~ \lambda^{2 \alpha _2} \qquad
{\rm and} \qquad s_{12} ~\sim~ \lambda^{\alpha _2} 
\ee

\begin{itemize}

\item First Solution. The central matrix is dominated by the lightest
right-handed neutrino mass $M_1$

\be
\frac{M_1}{M_3} ~\preceq~ s_{23}^2 \, \lambda^{2 \alpha _2} \qquad
{\rm and} \qquad s_{23} ~\succ~ \lambda^{\alpha _1} \, .
\ee
This leads to a large mixing angle $\phi _{12}$ in the 1-2 block
and small mixing angles with the third family
\be
\phi _{13} ~\sim~ \phi _{23} ~\sim~ 
\frac{\lambda^{\alpha _1}}{s_{23}} ~\prec~ 1
\ee
The physical neutrino masses are hierarchical, with
\be
m_{\nu _1} ~\sim~ m_{\nu _2} ~\sim~ 
\frac{\lambda^{2 \alpha _1}}{s_{23}^2} \, m_{\nu _3} ~\sim~ 
\frac{m^2}{M_1} \, \lambda^{2 \alpha _1 + 2 \alpha _2}   
\label{nu12}
\ee

\item Second Solution. We have the possibility of an inverted hierarchy 
between $M_2$ and $M_3$

\be
\lambda^{2\alpha _2} ~\preceq~ \frac{M_1}{M_3} ~\preceq~ 
\frac{\lambda^{2 \alpha _2}}{s_{23}^2}
\ee
This leads to
\be
\phi _{13} ~\prec~ \phi _{23} ~\sim~ 
s_{23} \, \lambda^{\alpha _1}
\ee
For the physical neutrino masses, $m_{\nu _1}$ and $m_{\nu _2}$
are still given by Eq.~(\ref{nu12}) but
\be
m_{\nu _3} ~\sim~ \frac{m^2}{M_3}   
\ee

\item Third Solution. The physical neutrino masses are the same as
for the second solution, but the mixing angles are different

\be
\max \{ \lambda^{2 \alpha _1 + 2 \alpha _2}\, , \,
s_{23}^2 \, \lambda^{2(\alpha _2 - \alpha _1)} \} ~\preceq~ 
\frac{M_1}{M_3} ~\preceq~ \lambda^{2 \alpha _2}
\ee
This gives
\be
\phi _{23} ~\sim~ \phi _{13} \, \lambda^{\alpha _1} ~\sim~ 
s_{23} \, \lambda^{\alpha _1 + 2 \alpha _2} \frac{M_3}{M_1}
\ee

\end{itemize} 

\item If the large angle is in the $1-3$ block of $\cal F$, the structure
of the central matrix obeys
\be
{\cal C}_{11} ~\sim~ {\cal C}_{13} ~\sim~ {\cal C}_{33} ~\prec~ {\cal C}_{22}
\ee
\be
{\cal C}_{12}^2\, , \,{\cal C}_{23}^2 ~\preceq~ 
{\cal C}_{11} \cdot~ {\cal C}_{22} 
\ee
We again get three different possibilities. To obtain the desired MNS matrix,
the heaviest neutrino must be labeled $\nu _3$, while the two lighter neutrinos
$\nu _1$ and $\nu _2$ mix with the large angle $\phi _{12}$ which gives rise to the
observed solar neutrino deficit. Following this labeling convention, we get

\begin{itemize}

\item First Solution. 
\be
\frac{M_2}{M_1} ~\preceq~ s_{12}^2 \, , \qquad
\frac{M_2}{M_3} ~\preceq~ s_{23}^2 \, , \qquad
s_{23} ~\sim~ s_{12} \, \lambda^{\alpha _1 + \alpha _2}
\ee
This gives 
\be
\phi _{13} ~\sim~ \phi _{23} ~\sim~ s_{12} \, \lambda^{\alpha _2}
\ee
and 
\be
m_{\nu _1} ~\sim~ m_{\nu _2} ~\sim~ 
\frac{m^2}{M_2} \, s_{12}^2 \, \lambda^{2 \alpha _1 + 2 \alpha _2} \, , \qquad
m_{\nu _3} ~\sim~ \frac{m^2}{M_2} \, \lambda^{2 \alpha _1}
\ee

\item Second Solution.

\be
\frac{M_1}{M_2} ~\preceq~ s_{12}^2 \, , \qquad
\lambda^{2 \alpha _1 + 2 \alpha _2} ~\preceq~ \frac{M_1}{M_3} ~\preceq~ 
\min \{ 1 \, , \, \frac{s_{12}^2}{s_{23}^2} \} \, , \qquad
s_{12} \, s_{23} ~\sim~ \lambda^{\alpha _1 + \alpha _2}
\ee
This gives
\be
\phi _{13} ~\sim~ \phi _{23} ~\sim~ s_{23} \, \lambda^{\alpha _1}
\ee
and
\be
m_{\nu _1} ~\sim~ m_{\nu _2} ~\sim~ 
\frac{m^2}{M_1} \, \lambda^{2 \alpha _1 + 2 \alpha _2} \, , \qquad
m_{\nu _3} ~\sim~ \frac{m^2}{M_1} \, s_{12}^2 
\ee

\item Third Solution. 

\be
\frac{M_1}{M_3} ~\preceq~ \lambda^{2 \alpha _1 + 2 \alpha _2} \, , \qquad
\frac{M_2}{M_3} ~\preceq~ s_{23}^2
\ee
and
\be
s_{12} \, s_{23} ~\sim~ \lambda^{\alpha _1 + \alpha _2} \, , \qquad
s_{12} ~\succ~ \lambda^{\alpha _2}
\ee
The mixing angles are now given by
\be
\phi _{13} ~\sim~ \phi _{23} ~\sim~ 
\frac{\lambda^{\alpha _2}}{s_{12}} ~\prec~ 1
\ee

\end{itemize} 

\item $\cal F$ has one large rotation in the $2-3$ block. Since this 
possibility cannot lead to the right MNS matrix, we will skip it.

\end{itemize}

We notice that it is possible to obtain one, three, or no  large 
angles in $\cal F$. To obtain only two large angles, there must 
be further fine-tuning conditions among the prefactors, or
partial cancellations in some matrix elements of the central 
matrix ${\cal C}$.
Without such partial cancellations, a generic solution cannot be obtained. 

\subsubsection{Case B}

We now turn to the possibility of partial cancellations in 
some matrix elements of ${\cal C}$, leading to a pseudo-Dirac spectrum
for the neutrinos which mix strongly ($\nu _1$ and $\nu _2$) and an
inverted hierarchy
\be
m_{\nu _1} ~\simeq~ -m_{\nu _2} ~\gg~ |m_{\nu _3}|
\ee
We notice that such spectrum cannot be obtained without partial cancellation.

\begin{itemize}

\item For ${\cal C}$ to have a pseudo-Dirac mass in the $1-2$ block, the only
condition is that the element ${\cal C}_{12}$ dominates over all the other
elements. There is only one solution, which extends the $(2 \times 2)$ result.
We have the partial cancellation
\be
\tan^2 \theta _{12} ~=~ -\frac{M_1}{M_2}
\left( 1+ {\cal O}(\lambda^\beta) \right)
\label{paca12}
\ee
together with
\be
\frac{M_1}{M_3} ~\prec~ s_{12} \, \lambda^{2 \alpha _1 + \alpha _2} \, , \qquad
\lambda^{\alpha _2} ~\prec~ s_{12} ~\prec~ \lambda^{\alpha _2 - \beta} \, , \qquad
s_{23} ~\prec~ \lambda^{\alpha _1} \, , \qquad
s_{13} ~\prec~ \lambda^{\alpha _1 + \alpha _2}
\ee
This gives
\be
\phi _{13} ~\sim~ \frac{s_{23}}{\lambda^{\alpha _1}} ~\prec~ 1 \, , \qquad
\phi _{23} ~\sim~ \frac{s_{13}}{\lambda^{\alpha _1 + \alpha _2}} ~\prec~ 1
\ee
and a pseudo-Dirac pair of absolute mass
\be
m_{\nu _1} ~\simeq~ -m_{\nu _2} ~\sim~ 
\frac{m^2}{\sqrt{-M_1 M_2}} \, \lambda^{2 \alpha _1 + \alpha _2}
\ee
The angle $\theta _{12}$ can be large while keeping $s_{23}$ and $s_{13}$
small as long as the partial cancellation condition Eq.~\ref{paca12} 
remains satisfied.

\item It is not possible to get a simple partial cancellation condition
that can give rise to a pseudo-Dirac mass in the $1-3$ block, if we keep the 
relation~(\ref{rel13}). This is expected since such pattern of the Majorana 
masses has an inverted hierarchy more consistent with the relations
\be
s_{23} ~\sim~ s_{12} \, s_{13} \qquad {\rm or} \qquad
s_{12} ~\sim~ s_{23} \, s_{13}
\label{newrel}
\ee
With either of these relations, we can indeed obtain a solution
with the partial cancellation condition
\be
\tan^2 \theta _{13} ~=~ -\frac{M_1}{M_3}
\left( 1+ {\cal O}(\lambda^\beta) \right)
\label{paca13}
\ee
For example, with the second relation in Eq.~(\ref{newrel}), we get
\be
\frac{M_1}{M_2} ~\prec~ \frac{s_{13}}{s_{23}^2} \, 
\lambda^{\alpha _1 + \alpha _2}\, , \, 
s_{13} \, \lambda^{\alpha _2 - \alpha _1} \, , \qquad
\lambda^{\alpha _1 + \alpha _2} ~\prec~ s_{13} ~\prec~ 
\lambda^{\alpha _1 + \alpha _2 - \beta} \, , \qquad
s_{23} \, s_{13} ~\prec~ \lambda^{\alpha _2}
\ee 
This gives
\be
\phi _{13} ~\sim~ \frac{s_{23}}{\lambda^{\alpha _1}} 
\max \{ 1\, , \, \frac{M_1}{M_2} \} ~\prec~ 1 \, , \qquad
\phi _{23} ~\sim~ \max \{ \frac{s_{23} \, s_{13}}{\lambda^{\alpha _2}} \, , \,
\frac{s_{23}}{s_{13} \, \lambda^{\alpha _2}} \frac{M_1}{M_2} \} ~\prec~ 1
\ee
and a pseudo-Dirac pair of absolute mass
\be
m_{\nu _1} ~\simeq~ -m_{\nu _2} ~\sim~ 
\frac{m^2}{\sqrt{-M_1 M_3}} \, \lambda^{\alpha _1 + \alpha _2}
\ee

\end{itemize}

Finally, let us notice that the structure Eq.~(\ref{invstruct}) which 
naturally leads to a bi-maximal mixing and an inverted hierarchy of
the light neutrinos, cannot be reproduced in the central matrix if
the relation~(\ref{rel13}) holds.

\section{Theoretical Outlook }

Having identified the conditions for two large mixing angles in the MNS matrix, 
it remains to be seen if they can be easily realized in credible theoretical schemes. 

Froggatt and Nielsen (FN) proposed long ago to parametrize the hierarchy in terms of 
effective operators coming from unknown interactions at higher energies, 
and that their degree of suppression is determined by extra charges; 
the higher the charge, the more suppressed the operator. 

This approach which explains hierarchies by adding $U(1)$ symmetries to the 
Standard Model has received much attention 
and shown some success especially when paired with anomalous Green-Schwarz 
$U(1)$ symmetries~\cite{GSM, ibanez, ross1, FM, me1, me2}.  

This is particularly suited to the Wolfenstein expansion as powers of the 
Cabibbo angle, as it relates the 
power of $\lambda$ to the FN charges of the basic quark and lepton fields. 
In the simplest form of the theory, these exponents can almost be read-off 
from the quark mass and mixing matrices.  
As an  example, after breaking the FN symmetry, the neutral lepton Dirac mass 
operator looks like

\be L^T_i\,{\cal M}^{(0)}_{ij}\,\overline N_j\ ,\ee
with the hierarchy determined by the charges of the lepton doublets $L_i$ and 
the right-handed neutral leptons $\overline N_i$.  
The matrix can be expunged from hierarchy by means of diagonal matrices

\be
{\cal M}^{(0)}_{}~=~\Lambda^{}_L\,\widehat{\cal M}^{(0)}_{}\,
\Lambda^{}_{\overline N}\ ,\ee
where 

\be
\Lambda~=~{\rm Diag}\,(\lambda_{}^{Q_1-Q_3}\ ,\lambda_{}^{ Q_2-Q_3}\ , 1\,)\ ,\ee
are the diagonal matrices and $Q_i$ are the FN charges of the relevant fields, 
and the hatted matrix contains only elements of order one. 

If the hierarchy is explained {\it solely} in terms of one FN field, one obtains 
a simple prediction for the seesaw masses, namely that the hierarchy matrix 
$\Lambda_{\overline N}$ coming from the right-handed neutrinos cancels 
out and that we are left with only the hierarchy coming from the lepton doublets. 
This conclusion does not hold in more complicated FN schemes with  fields of 
opposite FN charges, but we are assuming the simplest possibility here. 

In this case, the $\cal F$ matrix contains no elements suppressed 
by powers of the Cabibbo angle, and all its elements are of order one, the MNS matrix 
contains {\it three} large angles, and not two as indicated by the data. 
Hence  the simplest FN scheme has trouble explaining any correlated hierarchy, 
and typically does not produce a MNS matrix with two large mixing angles. 
To produce a correlated hierarchy in the FN approach, there must be several 
FN fields of charges of different signs.  
 
This seems to favor a textured Majorana matrix, with carefully chosen zeros, 
either through a particular texture or through supersymmetric zeroes.  
This leads to a simple proposal: use the same FN schemes proposed in our 
previous models for the $\Delta I_{\rm w}=1/2$ masses, but require that the   
$\Delta I_{\rm w}=0$ mass be textured with a Dirac mass for the right-handed 
neutrinos of the first two families, and a Majorana mass for the third, to 
zeroth order in the Cabibbo angle

\be 
\pmatrix{0&~M^{}_1~&0\cr ~M^{}_1~&0&0\cr 0&0&~M_3^{}~}~+~\cdots\ .\ee
Then as we have shown in section 5, this generates a nearly maximal angle 
between the first two neutrinos. 

The required texture for the $\Delta I_{\rm w}=0$ 
Majorana mass must be correlated with that of the $\Delta I_{\rm w}=1/2$ 
Dirac mass in such a way that the large angles coming from both sectors do 
not rotate the same block; this would produce only one large angle. 
In particular, if the charged lepton sector produces a large angle along 
$2-3$, the large angle in the central matrix must be either along $1-2$ or $1-3$, 
but not $2-3$. As argued throughout the paper, the presence of one small
mixing angle in the MNS matrix can be thought as the last remnant of a 
hierarchical structure in the lepton sector. 
Since hierarchical patterns are generic in the context of Grand 
Unified Theories and quark-lepton unification, our analysis shows that
it is much more natural to attribute the two observed large mixing angles
in the MNS matrix to different sectors~\cite{AB}. We conclude that the large atmospheric
angle must stem from the charged lepton sector, while the large solar angle
must reflect some peculiar behavior of the right-handed neutrino Majorana
sector and the seesaw mechanism.

\section{Acknowledgments}
This work is supported by the United States Department of Energy
under grant DE-FG02-97ER41029.


\begin{thebibliography}{99}

\bibitem{SKatm} The Super-Kamiokande Collaboration,
				\jrn{\PRL}{85}{3999}{2000}.	  
\bibitem{SKsol}	The Super-Kamiokande Collaboration,
				\jrn{\PLB}{539}{179}{2002}.	  
\bibitem{SNO} The SNO Collaboration,
			  \jrn{\PRL}{89}{011301}{2002}.
\bibitem{CHOOZ} M. Apollonio {\it et al.},
                \jrn{\PLB}{338}{383}{1998};
				\jrn{\PLB}{420}{397}{1998}.
\bibitem{kamland} The KamLAND Collaboration,
				  \jrn{\PRL}{90}{021802}{2003}.
\bibitem{K2K} The K2K Collaboration,
			  \jrn{\PRL}{90}{041801}{2003}.
\bibitem{SU5SO10} G. Altarelli, F. Feruglio and I. Masina,
			  	  \jrn{JHEP}{0011}{040}{2000};
			  	  W. Buchm\"uller,
				  Acta Phys.Polon. B32 (2001) 3707-3718;
				  T. Blazek, S. Raby, K. Tobe,
				  \jrn{\PRD}{62}{055001}{2000}.
\bibitem{SEESAW} M. Gell-Mann, P. Ramond, and R. Slansky in Sanibel Talk, 
				 CALT-68-709, Feb 1979 retropreprinted as hep-ph/9809459, 
				 and in {\it Supergravity} (North Holland, Amsterdam 1979). 
				 T. Yanagida, in {\it Proceedings of the Workshop on
				 Unified Theory and Baryon Number of the Universe}, 
				 KEK, Japan, Feb 1979.			  
\bibitem{HRR} J.A. Harvey, D.B. Reiss and P. Ramond,
			  \jrn{\NPB}{199}{223}{1982}.	
\bibitem{FM} P. Bin\'etruy, N. Irges, S. Lavignac and P. Ramond,
			 \jrn{\PLB}{403}{38}{1997};
			 N. Irges, S. Lavignac and P. Ramond,
		 	 \jrn{\PRD}{58}{035003}{1998}.	
\bibitem{FN} C. Froggatt and H.B. Nielsen,
			 \jrn{\NPB}{147}{277}{1979}.	
\bibitem{RRR} P. Ramond, R.G. Roberts and G.G. Ross,
			  \jrn{\NPB}{406}{19}{1993}.			 			  			
\bibitem{fogli} G.L. Fogli {\it et al.},
				\jrn{\PRD}{66}{093008}{2002};
				G.L. Fogli {\it et al.},
				\jrn{\PRD}{67}{073002}{2003}.
\bibitem{WMAP} The WMAP Collaboration,
			   astro-ph/0302207;	
			   astro-ph/0302209.									
\bibitem{KM} For an intriguing pattern, see P. Kaus and S. Meshkov, 
			 hep-ph/0211338.
\bibitem{me1} F.-S. Ling and P. Ramond,
			  \jrn{\PLB}{543}{29}{2002}.
\bibitem{sato} J. Sato and T. Yanagida,
			   \jrn{\PLB}{493}{356}{2000}.
\bibitem{vissani} F. Vissani,
				  \jrn{JHEP}{9811}{025}{1998};
				  F. Vissani,
				  \jrn{\PLB}{508}{79}{2001}.
\bibitem{me3} F.-S. Ling, hep-ph/0304135.

\bibitem{king} S.F. King,
			   \jrn{JHEP}{0209}{011}{2002}.
\bibitem{babu} K.S. Babu and R.N. Mohapatra
			   \jrn{\PLB}{532}{77}{2002}.
\bibitem{me2} F.-S. Ling and P. Ramond,	
			  \jrn{\PRD}{67}{115010}{2003}.		  			 		   			 			  		    

\bibitem{SYAB} J. Sato and T. Yanagida, 
			   \jrn{\PLB}{430}{127}{1998}; 
			   C. H. Albright, K. S. Babu and S. M. Barr 
			   \jrn{\PRL}{81}{1167}{1998};
			   C. H. Albright and S. M. Barr,
			   \jrn{\PRD}{58}{013002}{1998}.
				  			   				  
\bibitem{BR}  {\it Family Cloning} of the gauge group, 
			  by Mark J. Bowick and P. Ramond, 
			  \jrn{\PLB}{131}{367}{1983}.

\bibitem{SMIR} A similar analysis was done earlier by  
			   A. Yu. Smirnov, 
			   \jrn{\PRD}{48}{3264}{1993}.
	
\bibitem{GSM} M. Green and J. Schwarz,
			  \jrn{\PLB}{149}{117}{1984}.		
\bibitem{ibanez} L. Iba\~nez,
				 \jrn{\PLB}{303}{55}{1993}.		
\bibitem{ross1} L. Iba\~nez and G.G. Ross,
				\jrn{\PLB}{332}{100}{1994}.
\bibitem{AB} C. H. Albright and S. M. Barr, 
			 \jrn{\PLB}{461}{218}{1999}.

\end{thebibliography}
\end{document}